# Atomistic modeling of molecular beam epitaxy growth of $SrTiO_3$ and $Sr_2TiO_4$ thin films

Guangfu Luo[1,2,3,*] and Dane Morgan[1,†]

[1]Department of Materials Science and Engineering, University of Wisconsin-Madison, Wisconsin 53706, USA

[2]State Key Laboratory of Quantum Functional Materials, Department of Materials Science and Engineering, Southern University of Science and Technology, Shenzhen 518055, China

[3]Institute of Innovative Materials, Southern University of Science and Technology, Shenzhen 518055, China

[*] Contact author: luogf@sustech.edu.cn; †Contact author: ddmorgan@wisc.edu

## ABSTRACT

Molecular beam epitaxy (MBE) is renowned for its potential for atomic layer control, but unexpected growth mechanisms can potentially compromise this level of precision. In this study, we employ first-principles calculations to investigate the atomistic processes governing the MBE growth of perovskite $SrTiO_3$ and Ruddlesden-Popper $Sr_2TiO_4$ films on a $SrTiO_3$ substrate. We systematically explore the potential molecular species in the gas phase and their reactions and diffusion dynamics on the film surfaces and layer edges. Our analyses uncover three mechanisms of importance for understanding this type of growth. First, oxygen vacancies can be dynamically induced during surface diffusion on defect-free substrate and noticeably accelerate the diffusion processes. Second, while the SrO layer is expected to grow in a single-layer growth mode, the presence of potential $Ti_{Sr}$ defects may promote the formation of SrO islands. Lastly, adsorbed Ti atoms and $TiO_2$ molecules on the SrO bilayer can insert into SrO bilayers, resulting in an unexpected growth sequence. These findings may have broader implications for the MBE growth of metal oxide films and provide guidance for achieving improved control over their growth processes.

# I. INTRODUCTION

Transition metal (TM) oxides constitute a diverse family of compounds that manifest a wide range of scientifically and technologically important phenomena. The strongly correlated electrons in certain TM oxides give rise to properties like large topological Hall effects, high-$T_c$ superconductivity, and metal-insulator transition [1–7]. Some TM oxides exhibit high ionic conductivity, making them essential in solid oxide fuel cells [8] and batteries [9,10]. Previous experiments have also uncovered a high-mobility electron gas at the interface of two insulators $LaAlO_3$ and $SrTiO_3$ [11,12]. These intriguing properties make TM oxide thin films of great importance for applications in renewable energy, electronics, and fundamental physics.

Realizing those intriguing phenomena in TM oxide thin films often requires atomic-scale precision in material fabrication. For instance, the formation of a conducting $LaAlO_3$/$SrTiO_3$ interface occurs only when the interface is precisely grown as LaO/$TiO_2$; otherwise, it remains insulating [11,13]. Molecular beam epitaxy (MBE) stands as a state-of-the-art technique for oxide film growth, offering low impurities and enabling the sequential deposition of single atomic layers. However, unlike metals and typical semiconductors, many metal oxides exhibit low volatility, which poses a significant challenge to precise film growth [14]. Moreover, unexpected dynamic rearrangement can occur during the film growth process [15–17]. Therefore, theoretical simulations play a crucial role in understanding key growth processes, controlling interface sharpness, minimizing defects, and guiding the synthesis of materials.

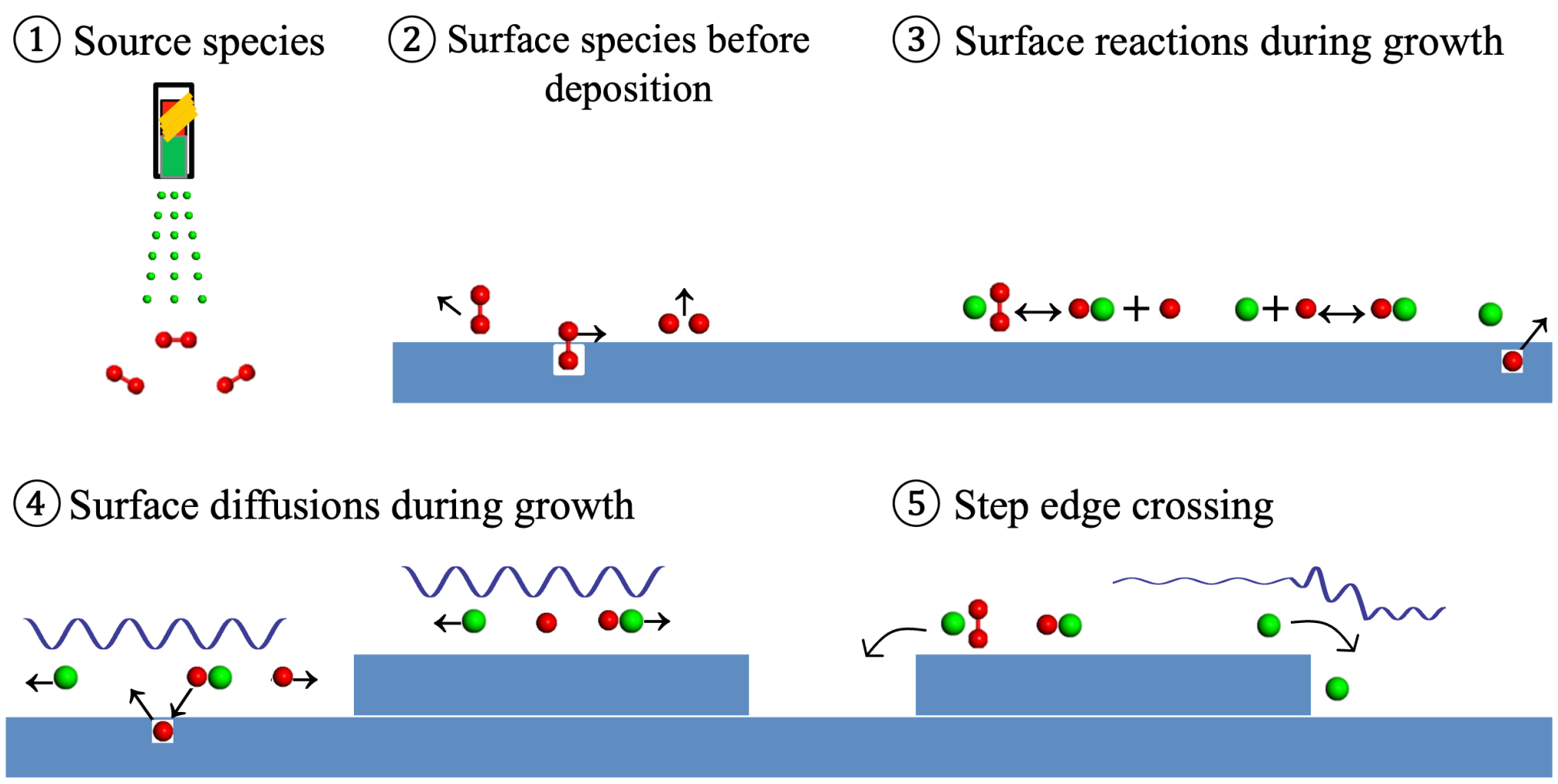

**FIG. 1.** Schematic processes during the MBE growth of $SrTiO_3$ and $Sr_2TiO_4$ films. The $SrO_x$ species are used for example.

In this study, we systematically investigate the MBE growth of $SrTiO_3$ and $Sr_2TiO_4$ on a $TiO_2$-terminated $SrTiO_3$ substrate using first-principles calculations. We explore processes in the gas phase, on film surfaces, and near layer edges, as schematically illustrated in Fig. 1. The thermodynamics and kinetics of adsorptions, hopping, and reactions are analyzed for species under relevant conditions. Finally, we compare our results with previous experimental findings to validate our first-principles calculations and provide insights into experiments.

## II. METHODS

All the first-principles thermodynamics, transition-state calculations, and *ab initio* molecular dynamics (AIMD) simulations are based on the density functional theory, as implemented in VASP [18]. The Perdew-Burke-Ernzerhof functional in the spin-polarized form is used for exchange-correlation functional [19]. A plane-wave energy cutoff of 500 eV is utilized, along with the following projector augmented-wave pseudopotentials [20]: $O(2s^2 2p^4)$ for O, $Ti(3s^2 3p^6 3d^2 4s^2)$ for Ti, and $Sr(4s^2 4p^6 5s^2)$ for Sr. All surfaces are modeled with (001) termination, consistent with the typical MBE growth on $SrTiO_3$ [11–13,15], and the in-plane lattice lengths and thickness are approximately 2.5 and 0.7 nm, respectively, with about 330 atoms per supercell. A vacuum layer of 1.5 nm thick is added in the out-of-plane direction to minimize interactions between periodic images. The Brillouin zone is sampled using the Γ point, and the total energy is converged with a tolerance of $10^{-5}$ eV per supercell. Transition states are identified using the climbing image nudged-elastic-band method [21], with 1–5 images placed between initial and final states, depending on the complexity of the dynamic process. Due to the substantial cancellation of thermal effects between the stable and transition states during surface hopping [22], no thermal corrections are considered in the relevant transition-state calculations. For selected cases, temperature-accelerated AIMD simulations are performed within the canonical ensemble at 2000 K and a time step of 1.5 fs to aid in the exploration of transition states. The timescale of a dynamic process is estimated using the activation barrier $E_a$ and the Arrhenius relation, as shown in Eq. (1):

$$t = v^{-1} e^{E_a / k_B T}, \tag{1}$$

where the attempt frequency $v$ is approximated as $5 \times 10^{12}$ s$^{-1}$. By comparing the timescale of an event with the growth time of each atomic layer, one can determine whether an event is fast under the growth conditions.

We adopt growth conditions based on previous MBE experiments [15]. In these experiments, $SrTiO_3$ and $Sr_2TiO_4$ films are grown on atomically flat $TiO_2$-terminated $SrTiO_3$ substrates using a sequential deposition mode, where the Sr and Ti beams are alternatively opened. The substrate temperature is 1023 K. The oxygen beam consists of $O_2$ gas with a partial pressure of $1.33 \times 10^{-4}$ Pa ($10^{-6}$ Torr), which corresponds to a flux rate of about $1.9 \times 10^{-2}$ molecules s$^{-1}$ Å$^{-2}$, as calculated using the ideal gas collision frequency on a wall, also known as the Hertz-Knudsen equation [23], as expressed in Eq. (2):

$$r_{flux} = \frac{P}{\sqrt{2\pi m k_{\mathrm{B}} T}}, \quad (2)$$

where $P$ and $m$ denote the partial pressure and molecular mass of a gaseous species, respectively. The Sr and Ti beams are generated from a Sr effusion cell and a Ti sublimation pump operating at 673 and 1773 K, respectively. While the partial pressures of the Sr and Ti beams were not directly measured, they are estimated to be $1.32 \times 10^{-2}$ and $3.09 \times 10^{-2}$ Pa, respectively, based on metal vapor pressure experiments [24]. The growth of each SrO or $TiO_2$ layer takes approximately 100 s, corresponding to a growth rate of $6.0 \times 10^{-4}$ cations s$^{-1}$ Å$^{-2}$. Note that this growth rate is much smaller than the flux rates calculated from Eq. (2), suggesting that a significant portion of the atoms in the Sr and Ti beams possibly rebound from the substrate.

To evaluate the desorption probability of a molecule $Sr_xTi_yO_z$ adsorbed on a surface, we define the adsorption free energy $G_{ad}$ as in Eq. (3):

$$G_{ad} = G(\mathrm{Sr}_x\mathrm{Ti}_y\mathrm{O}_z{*}) - G({*}) - x\mu(\mathrm{Sr}) - y\mu(\mathrm{Ti}) - 0.5z\mu(\mathrm{O}_2), \quad (3)$$

where $G(Sr_xTi_yO_z*)$, $G(*)$, $\mu(Sr)$, $\mu(Ti)$, and $\mu(O_2)$ represent the free energies of the surface with adsorbed molecule $Sr_xTi_yO_z$, a bare surface, Sr atom in the beam, Ti atom in the beam, and oxygen molecule in the beam, respectively. Here, the three chemical potentials are determined directly from the molecular beam conditions. These conditions differ from a previously proposed optimal growth window [25], where gaseous SrO exhibits an extremely low partial pressure and thus yields an impractically low growth rate. We speculate that the maximal SrO partial pressure predicted in that work, which was constrained by the thermodynamic limit for the condensation of gaseous SrO, can be exceeded due to kinetic barriers associated with the formation of solid SrO. Details of the growth conditions and the corresponding chemical potentials are summarized in Table S1 in the Supplemental Material [22].

Thermal corrections involve vibrational contributions and additional translational and rotational contributions for gaseous beams. For $G(Sr_xTi_yO_z*)$, thermal corrections are evaluated for both the adsorbed molecular species and slab atoms that are first-and second-nearest neighbors to the adsorbate, while keeping the remaining slab atoms fixed. This approach provides a computationally efficient yet reliable approximation for capturing the vibrational contribution to the adsorption free energy. The temperature of all species on the substrate is set to 1023 K, while the temperatures and partial pressures of gaseous species are 1023 K and $1.33 \times 10^{-4}$ Pa for $O_2$, 673 K and $1.32 \times 10^{-2}$ Pa for Sr, and 1773 K and $3.09 \times 10^{-2}$ Pa for Ti, respectively, consistent with the experimental conditions described earlier.

## III. RESULTS AND DISCUSSIONS

To investigate the MBE growth of $SrTiO_3$ and $Sr_2TiO_4$ films, we first examine the compositions of the molecular beams of $O_2$, Sr, and Ti. Considering the stacking sequence of SrO/$TiO_2$ in $SrTiO_3$ and SrO/SrO/$TiO_2$ in $Sr_2TiO_4$, we explore the corresponding diffusions and reactions involving $O_2^*$, $O^*$, $Sr^*$, $SrO^*$, and $SrO_2^*$ on the $TiO_2$ and SrO surfaces. Additionally, we analyze the Ehrlich-Schwoebel (E-S) barriers across the SrO step edge. Following that, we investigate the diffusions and reactions of $Ti^*$ and $TiO_2^*$ on SrO and bilayer SrO surfaces. Finally, we compare our predictions with available experimental findings. The optimized geometrical structures corresponding to figures in this study are provided in the Supplemental Material [22], and all the key input and output files are available through a third-party data repository [26].

### A. Compositions of the molecular beams

We begin with a thermodynamic analysis to determine the species likely to be evaporated and reach the substrate during MBE growth. Our results indicate that, under the given experimental conditions, the energy required to remove a Sr (Ti) atom from a Sr (Ti) surface is about 1.0 (1.1) eV lower than that of a $Sr_2$ ($Ti_2$) dimer. Furthermore, the dissociation reactions $Sr_2 \rightarrow 2Sr$ and $Ti_2 \rightarrow 2Ti$ are exothermic, with energy decreases of 0.39 and 1.05 eV per dimer, respectively. Therefore, the Sr and Ti beams are expected to primarily consist of single Sr and Ti atoms. Given that the atomization energy of $O_2$ is about 5.12 eV [27], $O_2$ is anticipated to remain as a dimer in the gas phase. Therefore, the three molecular beams are expected to consist primarily of $O_2$ molecules, Sr atoms, and Ti atoms.

### B. Growth of SrO layer on $TiO_2$-terminated surface

In the growth of both $SrTiO_3$ and $Sr_2TiO_4$, the formation of SrO layer on the $TiO_2$-terminated surface is essential. Because the oxygen beam remains open during the growth process, we first examine the interactions between the $O_2$ gas and the $TiO_2$-terminated $SrTiO_3$ substrate. Thermodynamic calculations indicate that the adsorption of $O_2$ on the $TiO_2$ surface is endothermic (Table I), largely because of the low partial pressure and high growth temperature. Consequently, $O_2$ is likely to collide and rebound from a perfect $TiO_2$ surface, a conclusion supported by our AIMD simulations at 1023 K. Given the likely presence of oxygen vacancy ($V_O$) in the $TiO_2$-terminated surface, we further investigate the reaction of $O_2$ with $V_O$. It is found that an $O_2$ molecule can directly occupy a $V_O$ site without breaking the O = O bond, leaving one

O atom above and forming a new O–Ti bond on the surface [structure I in Fig. 2(a)]. This process occurs spontaneously without any barrier [24]. This O* can hop to a neighboring site I after overcoming an energy barrier of 0.22 eV [Fig. 2(d)]. A less stable site for O* is located above a Ti atom [structure II in Fig. 2(a)], which is 0.29 eV higher and separated by a forward energy barrier of 0.75 eV from site I [Fig. 2(d)]. The two energy barriers are consistent with previous results of 0.30 and 0.68 eV for oxygen adatom hopping [28]. Therefore, $O_2$ gas can easily annihilate the $V_O$ defects in the $TiO_2$-terminated surface, and the resulting O* atoms diffuse rapidly on the surface.

When two O* atoms come into proximity on the $TiO_2$-terminated surface, they may form an $O_2$ molecule and desorb. As illustrated in Fig. 2(b), two neighboring O* atoms preferentially form in configuration IV, which is 0.43 and 0.84 eV lower than configurations III [Fig. 2(b)] and V [Fig. 2(c)], respectively. The $O_2$ desorption through configuration IV exhibits a barrier of 1.73 eV [Fig. 2(e)] relative to the isolated O* state, lower than that through structure V [Fig. 2(f)]. According to Eq. (1), the $O_2$ desorption occurs on a timescale of 67 μs at the growth temperature of 1023 K. Further calculations of $V_O$ formation energy (∼0.61 and ∼2.87 eV on the $TiO_2$- and SrO-terminated surfaces, respectively, under the growth temperature and oxygen partial pressure) indicate that the equilibrium fractional concentration of $V_O$ on the $TiO_2$-terminated surface is on the order of $e^{-0.61eV/(1023\ k_B)} \approx 10^{-3}$. This implies that roughly 1 out of every 1000 surface oxygen atoms tends to form a $V_O$.

**TABLE I.** Adsorption free energies of $O_2$, Sr, and Ti on the $TiO_2$-, SrO-, and bilayer SrO-terminated surfaces based on the beam pressures and temperatures in experiments (see Methods). All energy units are in eV.

| Molecule | $TiO_2$ termination | SrO termination | Bilayer SrO termination |
|---|---|---|---|
| $O_2$ | 3.05 | 2.92 | 3.02 |
| Sr | -2.67 | 0.24 | 0.07 |
| $SrO_2$ | -1.90 | 0.02 | – |
| SrO | -4.02 | -1.18 | – |
| Ti[a] | – | 1.15 | 0.36 |
| $TiO_2$ | – | -3.76 | -5.96 |

[a]Adsorbed Ti can rotate on the SrO- and bilayer SrO-terminated surface, and the results at the initial adsorption sites are shown here.

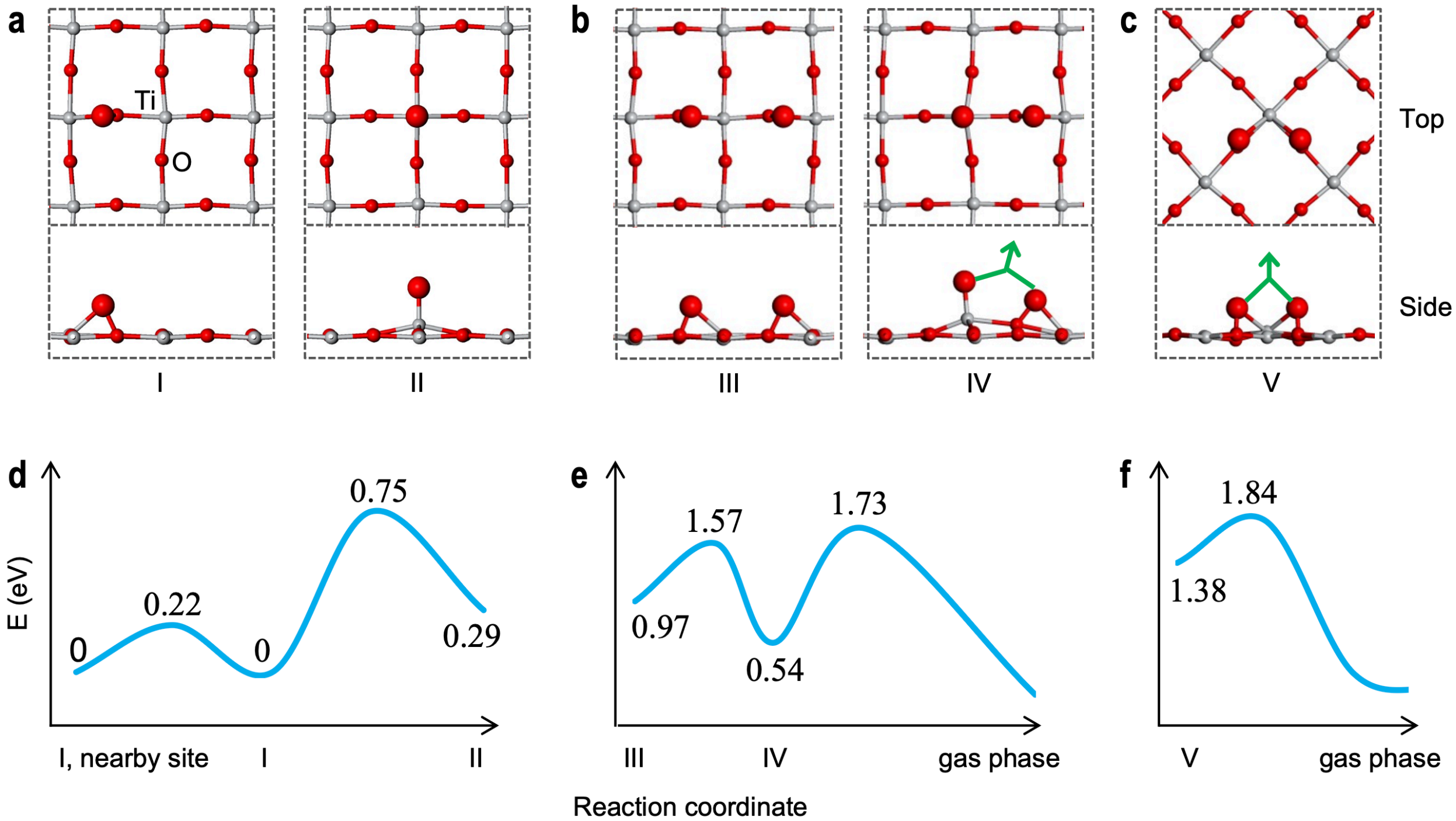

**FIG. 2.** (a) $O_2$ filling a $V_O$ in the $TiO_2$-terminated surface to form two O* states; (b) two neighboring $O^*$ atoms leading to the formation and desorption of $O_{2(g)}$; and (c) an alternative path leading to the formation and desorption of $O_{2(g)}$. Structures below the top layer are hidden for clarity. (d)–(f) Energy landscapes of processes related to the states in panels (a)–(c). To facilitate the comparison among different states, the energy of isolated O* is set as the energy reference in panels (d)–(f).

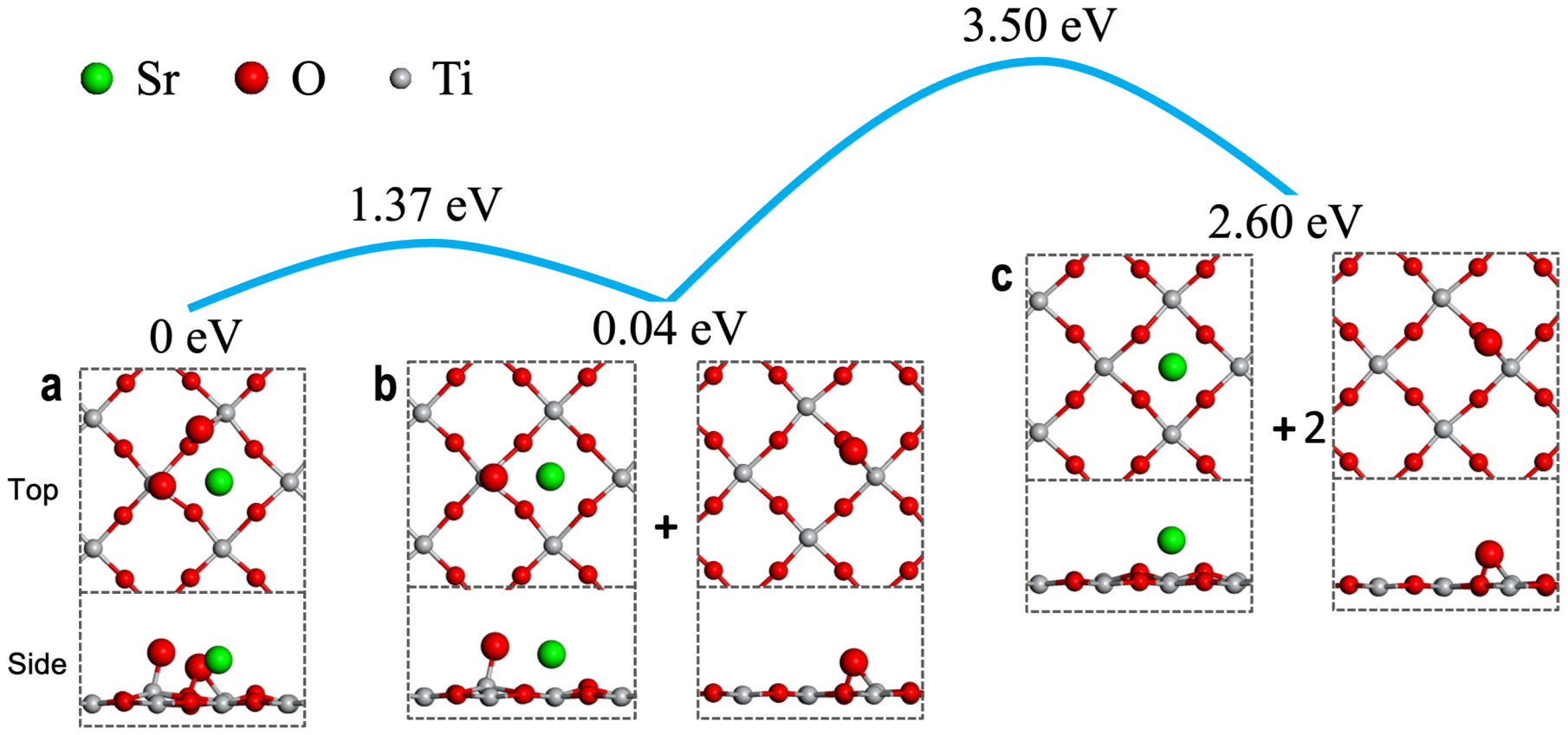

**FIG. 3.** Structures and relative energetics in the processes of $SrO_2^* \leftrightarrow SrO^* + O^* \leftrightarrow Sr^* + 2O^*$ on the $TiO_2$-terminated surface. Values above each structure indicate the energy relative to the lowest one in the figure.

With the above understanding, we now focus on the processes involving Sr monomer and dimer on the $TiO_2$-terminated surface, with four primary reactions: $Sr^* + O_{2(g)} \rightarrow SrO_2^*$, $SrO_2^* \leftrightarrow SrO^* + O^*$, $SrO^* \leftrightarrow Sr^* + O^*$, and $Sr^* + Sr^* \leftrightarrow 2Sr^*$. As shown in Fig. 3, $Sr^*$ resides above the square center of the $TiO_2$ layer, with a strong adsorption energy of −2.67 eV (Table I). It readily reacts with $O_2(g)$ to form $SrO_2^*$ without an energy barrier. The subsequent decomposition of $SrO_2^* \rightarrow SrO^* + O^*$ occurs with an energy barrier of 1.37 eV, and the final state is only 0.04 eV higher (Fig. 3). Further decomposition of $SrO^* \rightarrow Sr^* + O^*$ is rare due to a substantial barrier of 3.46 eV. Conversely, the formation reaction of $Sr^* + O^* \rightarrow SrO^*$ is relatively facile, with an energy barrier of 0.90 eV and a final state 2.56 eV lower. The formation of Sr dimer through the reaction $Sr^* + Sr^* \rightarrow 2Sr^*$ is energetically unfavorable, as the final state is 0.79 eV higher. Consequently, $Sr^*$, $SrO_2^*$, $SrO^*$, and $O^*$ are anticipated to be the dominant species during the initial growth of SrO layer on the $TiO_2$-terminated surface.

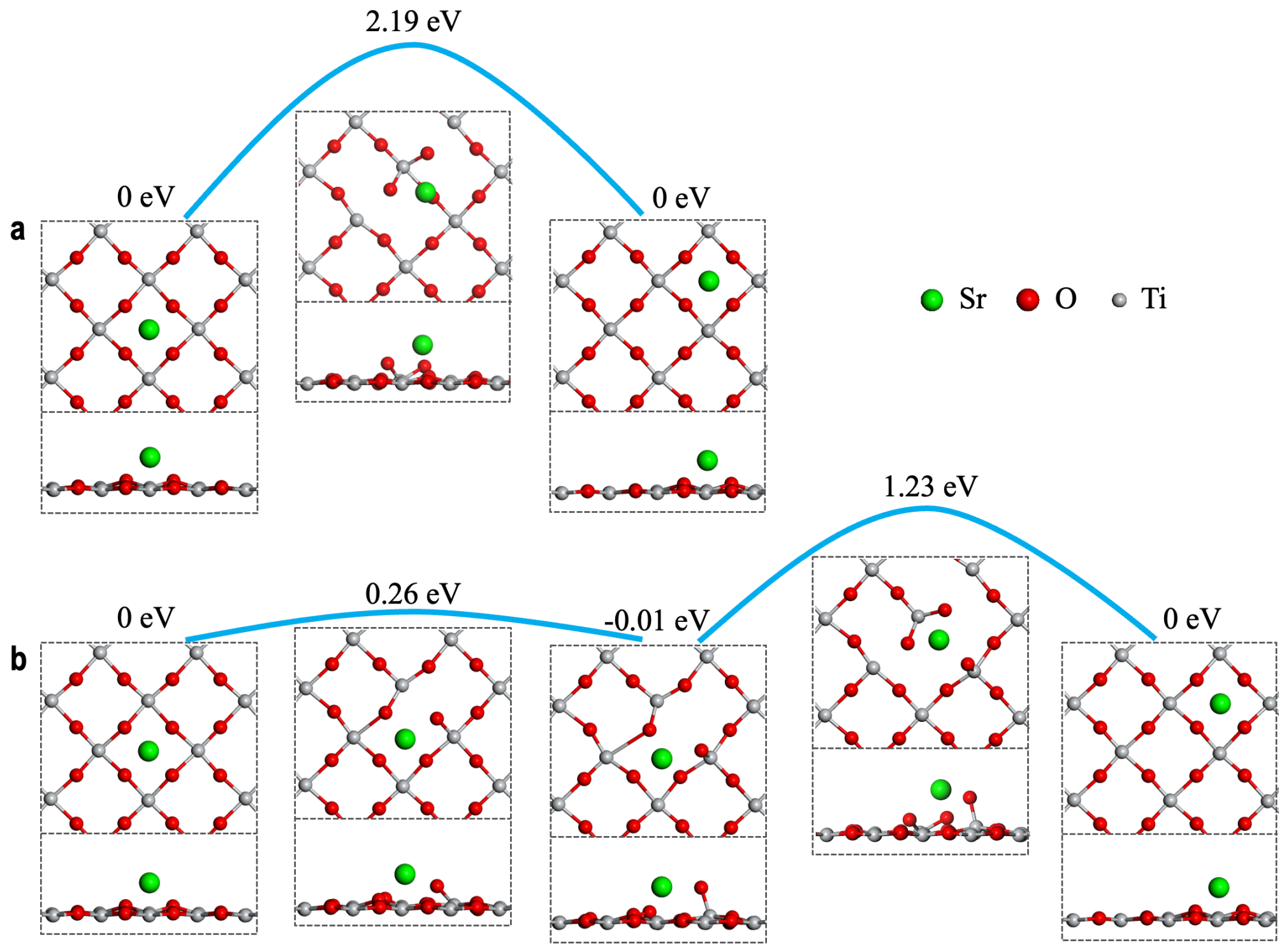

**FIG. 4.** Hopping of $Sr^*$ on the $TiO_2$-terminated surface (a) directly from one square center to a neighboring one or (b) through the generation of a $V_O$ nearby.

We next examine the diffusions of $Sr^*$, $SrO^*$, and $SrO_2^*$ on the $TiO_2$-terminated surface [the $O^*$ diffusion is presented in Fig. 2(d)]. Figure 4(a) illustrates the direct hopping path of $Sr^*$ from one square center to a neighboring one, a process with an energy barrier of 2.19 eV. To ascertain the dominance of this path, we conduct an AIMD simulation, which unexpectedly reveals that $Sr^*$ pulls a nearby O atom out of the $TiO_2$ plane and consequently creates a $V_O$. Further calculation reveals that this new state lies 0.01 eV below the initial state and is separated from it by an energy barrier of 0.26 eV [Fig. 4(b)]. For comparison, transferring an oxygen atom from a perfect $TiO_2$-terminated surface to a site analogous to that involved in the Sr* hopping requires an energy increase of 1.13 eV [22], indicating that Sr* can substantially facilitate the formation of oxygen vacancy. When $Sr^*$ hops via the $V_O$, a faster path with an energy barrier of 1.23 eV is established, which corresponds to a hopping timescale of 0.2 μs.

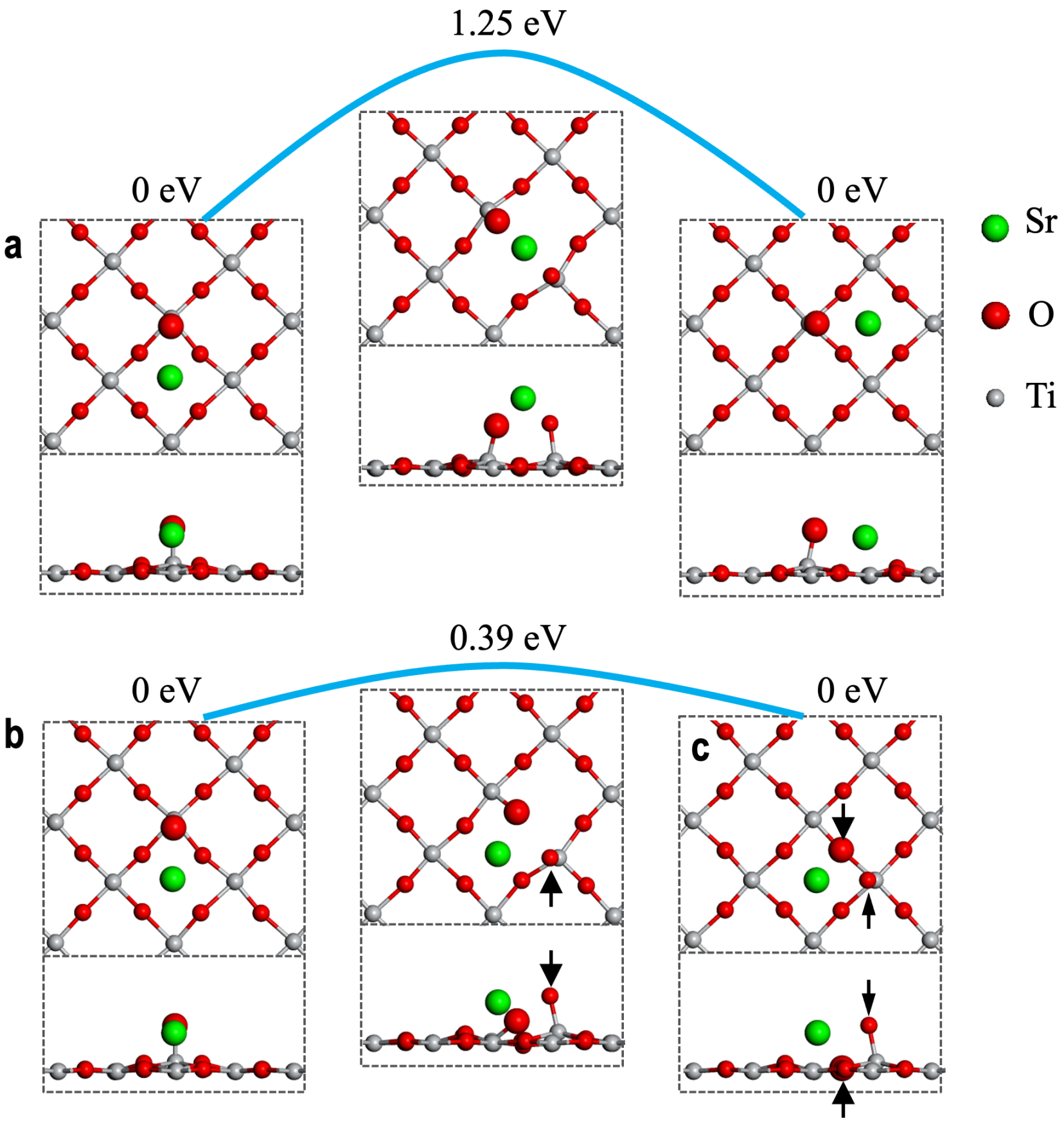

**FIG. 5.** Hopping of $SrO^*$ on the $TiO_2$-terminated surface for (a) O and (b) Sr atom, respectively. Certain oxygen atoms are indicated by black arrows for clarity.

Regarding $SrO^*$, its most stable adsorption position places the Sr atom above the $TiO_2$ square center and the O atom above a Ti atom [Fig. 5(a)]. Direct hopping of the Sr or O atom in $SrO^*$ to an equivalent neighboring position exhibits a significant energy barrier of 1.59 or 2.90 eV, the latter of which correspond to a timescale of ~39 s, the same timescale to grow a complete atomic layer. To resolve this contradiction, we conduct AIMD simulations, which reveal that a $V_O$ in the $TiO_2$ plane can be easily induced near the $SrO^*$, similar to the case for $Sr^*$ [Fig. 4(b)]. Therefore, we propose two new hopping mechanisms involving this $V_O$. For the Sr atom of $SrO^*$, it hops by jumping over a temporarily formed $V_O$ in the $TiO_2$ surface [Fig. 5(a)], similar to $Sr^*$ [Fig. 4(b)]. For the O atom of $SrO^*$, it hops by exchanging an O in the $TiO_2$ surface, effectively pulling one O atom out from the surface and then filling in the $V_O$ [Fig. 5(b)]. Subsequent calculations confirm that the two new mechanisms exhibit significantly lower energy barriers: 1.25 eV for the hopping of Sr and 0.39 eV for the hopping of O. Consequently, the timescale of SrO* hopping is reduced to 0.3 μs, much faster than the growth time of each atomic layer. These unexpected hopping paths underscore the importance of considering some slightly excited states in transition-state calculations, where AIMD simulations can offer crucial insights.

In contrast with the relatively fast diffusion of $Sr^*$ and $SrO^*$, $SrO_2^*$ is unable to diffuse as a whole on the $TiO_2$ surface. This is due to a significant energy increase (about 3 eV) when the Sr atom of $SrO_2^*$ is moved to a neighboring site. Given this substantial energy cost, no further hopping barrier of $SrO_2^*$ is considered. Nevertheless, as shown in Fig. 3, $SrO_2^*$ can decompose into $SrO^*$ and $O^*$ after overcoming an energy barrier of 1.37 eV. Subsequently, $SrO^*$ and $O^*$ can diffuse on the surface.

**TABLE II.** Hopping barriers of $O^*$, $Sr^*$, $SrO^*$, and $SrO_2^*$ on the $TiO_2$- or SrO-terminated surface, and across the SrO [100] edge. All values are in the energy unit of eV. Values in parentheses are results without utilizing $V_O$ in the $TiO_2$ plane. Subscripts indicate the hopping atoms of each polyatomic molecule.

| Species | $TiO_2$ termination | SrO termination | SrO [100] edge |
|---|---|---|---|
| $O^*$ | 0.22, 0.75 | 0.21, 0.80 | 0.83 |
| $Sr^*$ | 1.23 (2.19) | 0.88 | 0.24 |
| $SrO^*$ | $1.25_{Sr}$ ($1.59_{Sr}$), $0.39_{O}$ ($2.90_{O}$) | $0.54_{Sr}$, $0.47_{O}$ | 0.28 |
| $SrO_2^*$ | Decompose into $SrO^*$ and $O^*$ and diffuse | $0.69_{Sr}$, $0.18_{O2}$ | 0.27 |

We summarize the hopping barriers for $O^*$, $Sr^*$, $SrO^*$, and $SrO_2^*$ on the $TiO_2$-terminated surface, along with other relevant cases, in Table II. The results indicate that all these species exhibit rapid diffusion on the $TiO_2$-terminated surface, with the maximum hopping barrier not exceeding 1.25 eV. At the growth temperature of 1023 K, this corresponds to a timescale of about 0.3 μs per hop, allowing extensive diffusion necessary for layer-by-layer growth.

### C. Growth of SrO layer on SrO-terminated surface

To understand the formation of double SrO layer during the growth of $Sr_2TiO_4$, we investigate the growth of SrO layer on the SrO-terminated surface. Similar to reactions on the $TiO_2$-terminated surface, we examine four reactions on the SrO-terminated surface: $Sr^* + O_{2(g)} \rightarrow SrO_2^*$, $Sr^* + O_2^* \rightarrow SrO_2^*$, $SrO_2^* \leftrightarrow SrO^* + O^*$, $SrO^* \leftrightarrow Sr^* + O^*$, and $Sr^* + Sr^* \leftrightarrow 2Sr^*$ (Fig. 6). It is found that $O_{2(g)}$ can attach to $Sr^*$ on the SrO surface to form $SrO_2$* without an energy barrier and the free energy is reduced by 0.22 eV (Table I). Also, the reaction of $Sr^* + O_2^* \rightarrow SrO_2^*$ is energetically favorable, reducing the total energy by 3.48 eV [Figs. 6(a) and 6(b)]. The $SrO_2^*$ forms an isosceles triangle in the (001) plane, with $Sr^*$ positioned above an O atom and $O_2^*$ positioned above two Sr atoms of the SrO surface [Fig. 6(b)]. The decomposition of $SrO_2^*$ → $SrO^*$ + $O^*$ on the SrO surface possesses an energy barrier of 1.16 eV and increases the total energy by 0.40 eV [Figs. 6(b) and 6(c)]. Interestingly, $SrO^*$ does not exhibit a rock-salt stacking on the SrO-terminated surface. Instead, it pulls one O atom out of the surface to form an isosceles triangle of $SrO_2$ in the $(1\bar{1}0)$ plane, leaving a $V_O$ beneath the Sr of $SrO_2$ [Fig. 6(c)]. Further decomposition of $SrO^*$ → $Sr^*$ + $O^*$ increases the energy to 3.68 eV, making this step unlikely to occur [Fig. 6(d)]. Therefore, $Sr^*$, $SrO_2^*$, $SrO^*$, and $O^*$ are expected to be the major species on the SrO-terminated surface during growth.

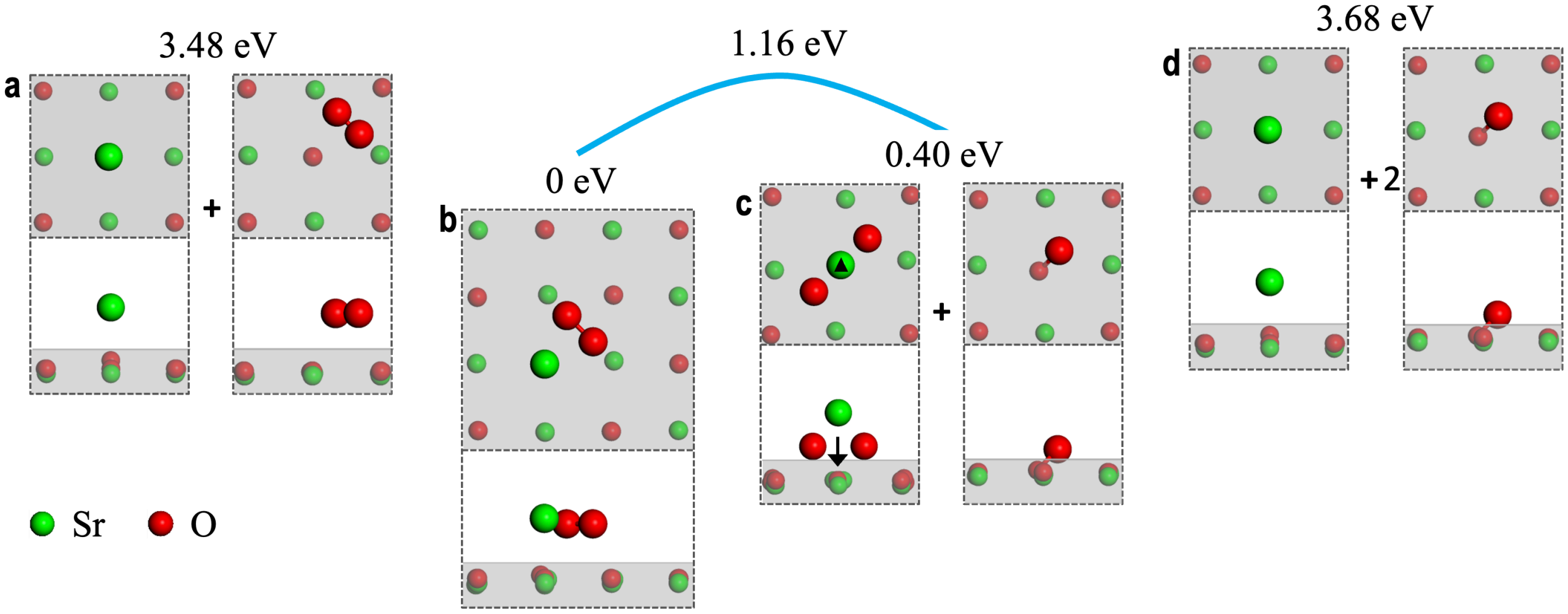

**FIG. 6.** Structures and relative total energies of the processes $Sr^*$ + $O_2^*$ → $SrO_2^*$ ↔ $SrO^*$ + $O^*$ ↔ $Sr^*$ + $2O^*$ on the SrO-terminated surface. In panel (c), an oxygen vacancy right below a $Sr^*$ is indicated by triangle and arrow in the top and side views, respectively. The SrO layers in this and following figures are shaded in gray for clarity.

Next, we examine the diffusion of $O^*$, $Sr^*$, $SrO^*$, and $SrO_2^*$ on the SrO-terminated surface. $O^*$ exhibits two types of hopping. First, it can hop to a neighboring equivalent position along the [110] direction, characterized by two symmetric transition states with a rate-limiting barrier of 0.80 eV and a minor barrier of 0.16 eV [Fig. 7(a) depicts half of the path]. Alternatively, it can rotate within the plane by 90°, presenting a barrier of 0.21 eV. Both results align with previously reported values of 0.81 and 0.20 eV, respectively [28]. Figure 7(b) illustrates that $Sr^*$ can hop to a neighboring equivalent site over a single barrier of 0.88 eV.

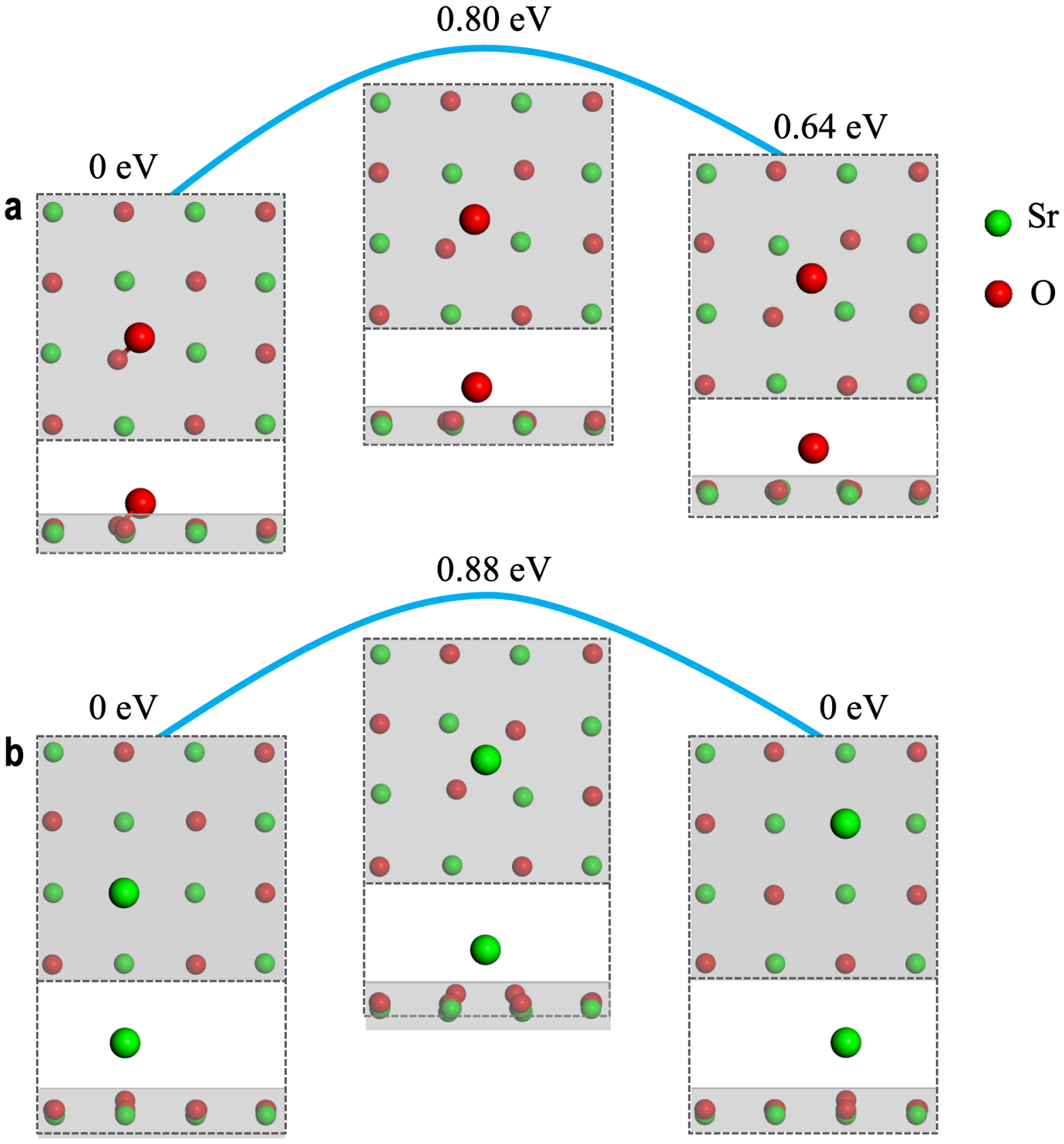

**FIG. 7.** Hopping of (a) $O^*$ and (b) $Sr^*$ on the SrO-terminated surface. Due to the symmetry of $O^*$ hopping path, only half of the path is shown in panel (a).

For $SrO^*$, the O rotation and Sr hopping mechanisms are depicted in Figs. 8(a)–8(e) and 8(a), 8(f)–8(i), respectively, both involving the filling and recreation of a $V_O$ in the SrO surface. The corresponding energy barriers for the O rotation and Sr hopping are 0.47 and 0.54 eV. In the case of $SrO_2^*$, the rotation of $O_2$ is notably easy, with an energy barrier of 0.18 eV [Figs. 8(j)–8(n)], while the Sr hopping possesses an energy barrier of 0.69 eV [Figs. 8(j), 8(o)–8(r)].

As summarized in Table II, $Sr^*$, $SrO^*$, and $SrO_2^*$ diffuse faster on the SrO-terminated surface than on the $TiO_2$-terminated surface, with the maximum energy barrier no greater than 0.88 eV. At the growth temperature of 1023 K, this corresponds to a timescale of ~2 ns per event, which is fast enough relative to the growth time (~100 s per atomic layer) to support substantial diffusion for layer-by-layer growth.

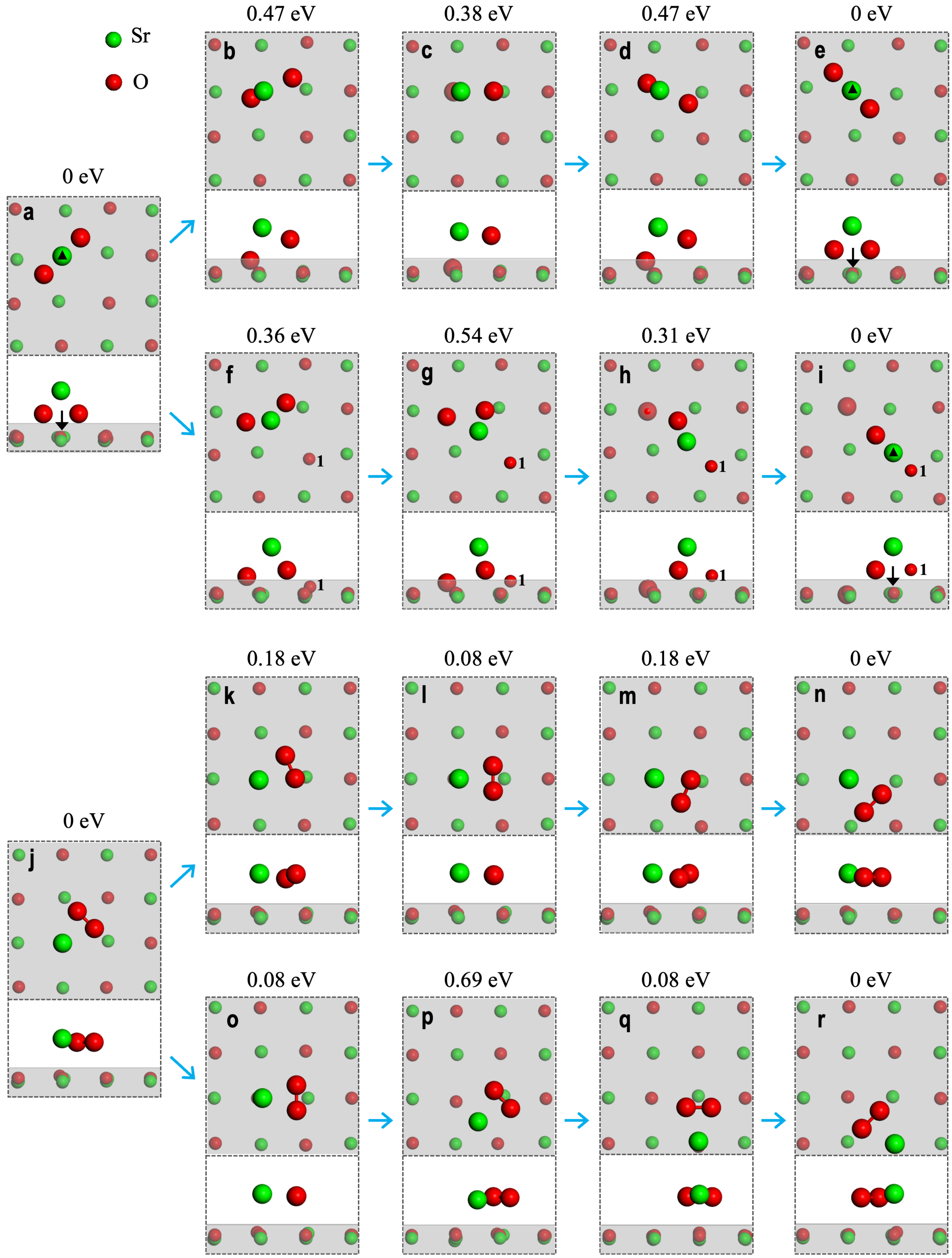

**FIG. 8.** Hopping of (a)–(i) $SrO^*$ and (j)–(r) $SrO_2^*$ on the SrO-terminated surface. In panels (a), (e), and (i), a $V_O$ below a $Sr^*$ is indicated by triangle and arrow in the top and side views, respectively. An oxygen atom in panels (f)–(i) is labeled for clarity.

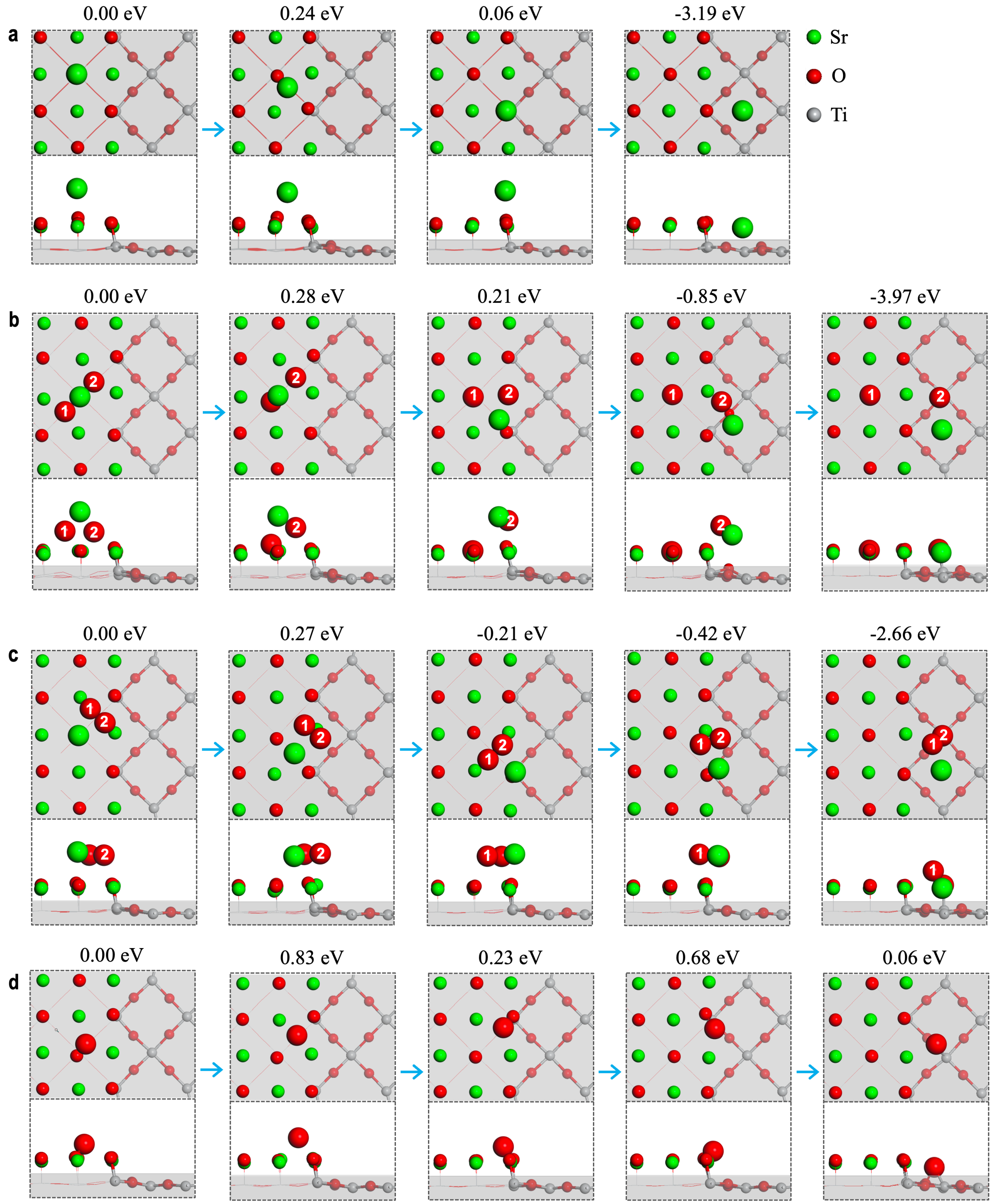

**FIG. 9.** Hopping of (a) $Sr^*$, (b) $SrO^*$, (c) $SrO_2{}^*$, and (d) $O^*$ across the SrO [100] edge. Atoms are shown in different styles, and some are numbered for clarity.

### D. Growth of SrO near the step edge of SrO[100]

The E-S barrier, an elevated energy obstacle relative to the in-plane diffusion barrier when an adsorbed species moves down a step edge [29], is crucial for understanding film growth. If the E-S barrier is sufficiently large, most species deposited on an existing island will remain there, leading to three-dimensional island growth. Conversely, a low E-S barrier promotes two-dimensional (2D) layer growth. Thus, we examine the hopping barriers of $Sr^*$, $SrO^*$, $SrO_2^*$, and $O^*$ across the most stable SrO [100] edge. As depicted in Figs. 9(a)–9(c), the edge-hopping barriers for $Sr^*$, $SrO^*$, and $SrO_2^*$ are 0.24, 0.28, and 0.27 eV, respectively, lower than the corresponding diffusion barriers of 0.88, 0.54, and 0.69 eV on the SrO-terminated surface (Table II). Therefore, no E-S barrier exist for these species. For O*, the edge-hopping barrier is 0.83 eV, resulting in a negligible E-S barrier of 0.03 eV compared with the diffusion barrier of 0.80 eV in the SrO-terminated surface (Table II). A summary of all the rate-limiting energy barriers related to the SrO layer growth on the $TiO_2$- and SrO-terminated surfaces, as well as those near step edges, is provided in Fig. 10. As a result, the SrO layer is expected to grow in a 2D growth mode.

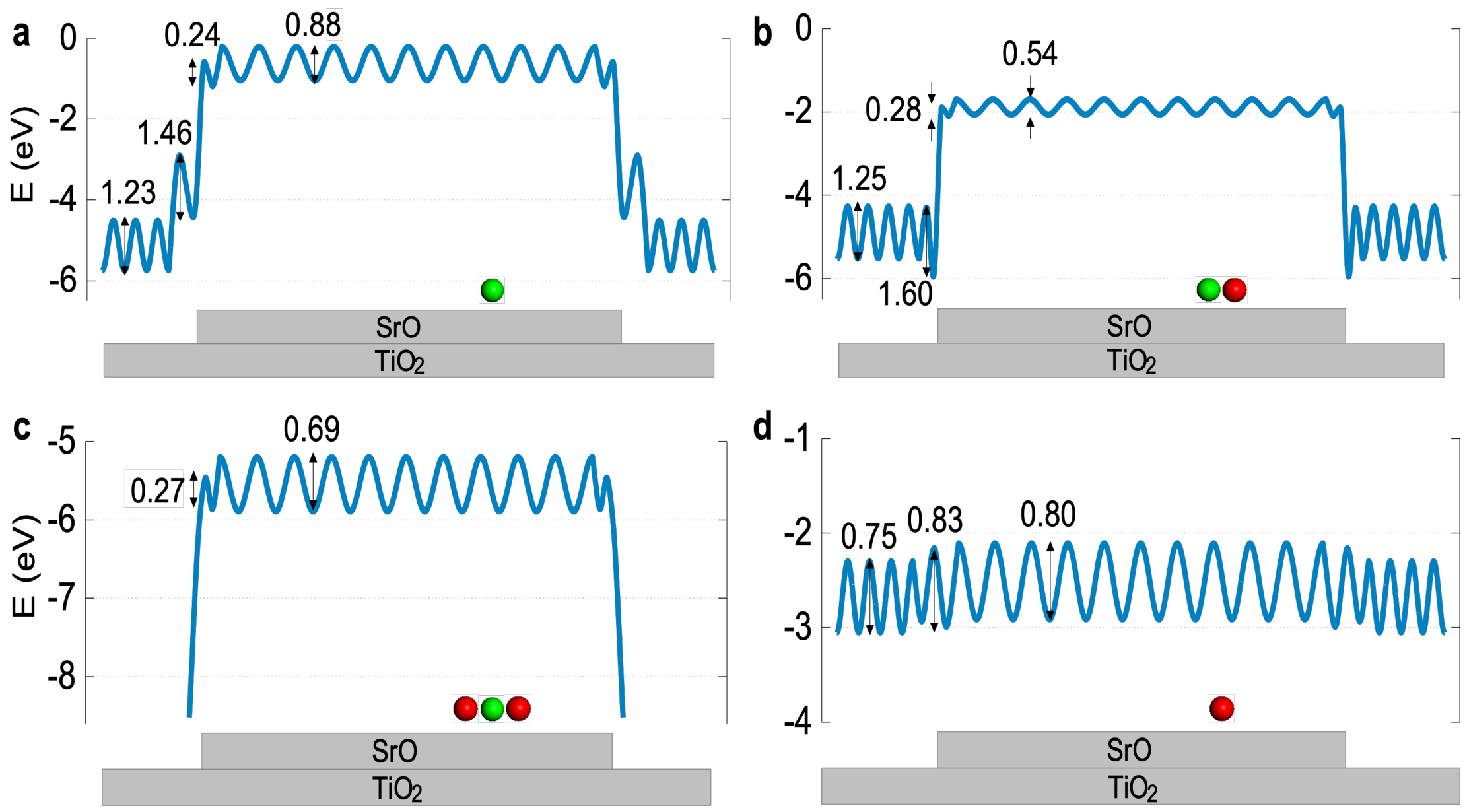

**FIG. 10.** Summary of hopping barriers of (a) $Sr^*$, (b) $SrO^*$, (c) $SrO_2^*$, and (d) $O^*$ on the $TiO_2$- and SrO-terminated surfaces and across the SrO[100] edge. Because $SrO_2^*$ tends to decompose into $SrO^*$ and $O^*$ on the $TiO_2$-terminated surface, no corresponding diffusion is shown in panel (c). For a diffusion process involving more than one barrier, only the rate-limiting one is provided.

We notice that previous experiments [30] on the epitaxial growth of $SrTiO_3$ have also observed the formation of multilayer SrO islands instead of a monolayer. This behavior was attributed to potential defects, such as Ti insertion into the SrO layer [30]. In Fig. 11, we present the structures and binding energies of $O^*$, $O_2^*$, $Sr^*$, $SrO^*$, and $SrO_2^*$ in association with a substitutional defect $Ti_{Sr}$ in the SrO-terminated surface. The results show that the $Ti_{Sr}$ defect exhibits very strong binding with $O^*$, $O_2^*$, $SrO^*$, and $SrO_2^*$, leading to energy decreases of 4.45, 3.88, 2.48, and 1.64 eV, respectively. These pronounced binding energies highlight the ability of a $Ti_{Sr}$ defect to anchor these species, thereby acting as the nucleus for SrO island formation. In addition to defects, authors of previous studies have indicated that strain may also drive a transition from layer-by-layer growth to island growth as a means of strain relaxation [31,32].

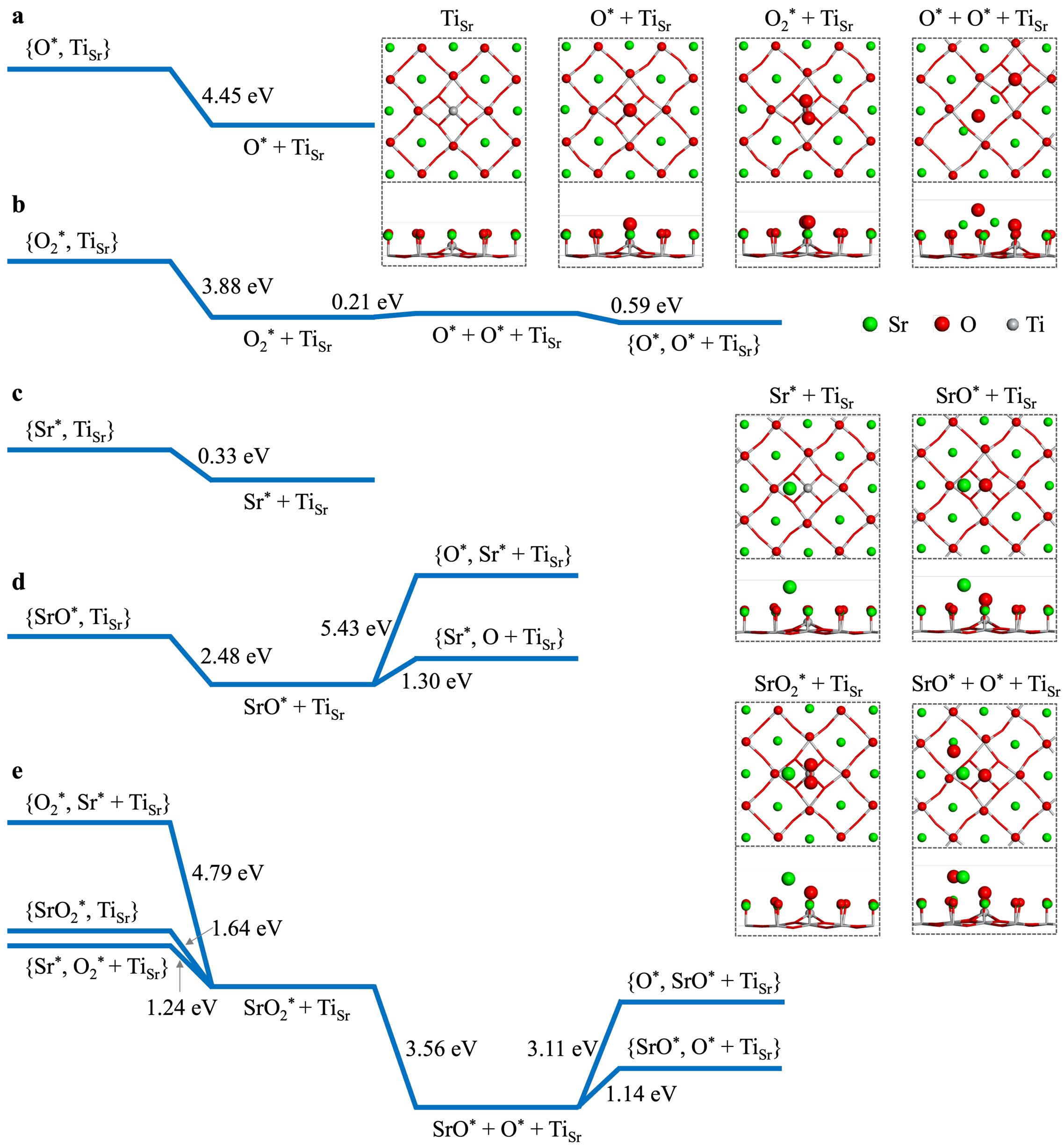

**FIG. 11.** Structures and relative energies of (a) $O^*$, (b) $O_2^*$, (c) $Sr^*$, (d) $SrO^*$, and (e) $SrO_2^*$ interacting with a $Ti_{Sr}$ defect in the SrO-terminated surface. The $TiO_2$ layer below surface is also shown, as the $Ti_{Sr}$ defect forms bonds with it. {*A*, *B*} denotes that *A* and *B* are isolated, while *A* + *B* denotes their proximity. The value beside each slope represents the energy difference of two relevant states.

### E. Growth of $TiO_2$ layer on SrO-and bilayer SrO-terminated surfaces

Figure 12(a) illustrates that $Ti^*$ on the SrO-terminated surface binds with two O atoms in the SrO surface to form a $TiO_2^*$ molecule. One oxygen in the $TiO_2^*$ molecule tilts upward and thus induces a $V_O$ in the surface. The $TiO_2^*$ molecule can rotate on the surface, transitioning to a $TiO_3^*$ molecule [Fig. 12(b)] and eventually to a $TiO_4^*$ molecule [Fig. 12(c)], with energy increases of 0.44 and 1.19 eV, respectively. The energy barriers for the $TiO_2^* \rightarrow TiO_3^*$ and $TiO_3^* \rightarrow TiO_4^*$ transitions are 1.44 and 1.31 eV, respectively. Additionally, the $TiO_2^*$ molecule can react with $O_2$ in the gas phase to form a $TiO_4^*$ molecule [Fig. 12(d)] without an energy barrier. Once the $TiO_4^*$ molecule is formed, it becomes largely immobile because of its strong bonding with the surface. This reduced mobility is anticipated to hinder island formation, akin to the effects of lowering the growth temperature or increasing the growth rate [33]. At the high growth temperature of 1023 K, these $TiO_x$ molecules are expected to coalesce and eventually form a continuous $TiO_2$ layer.

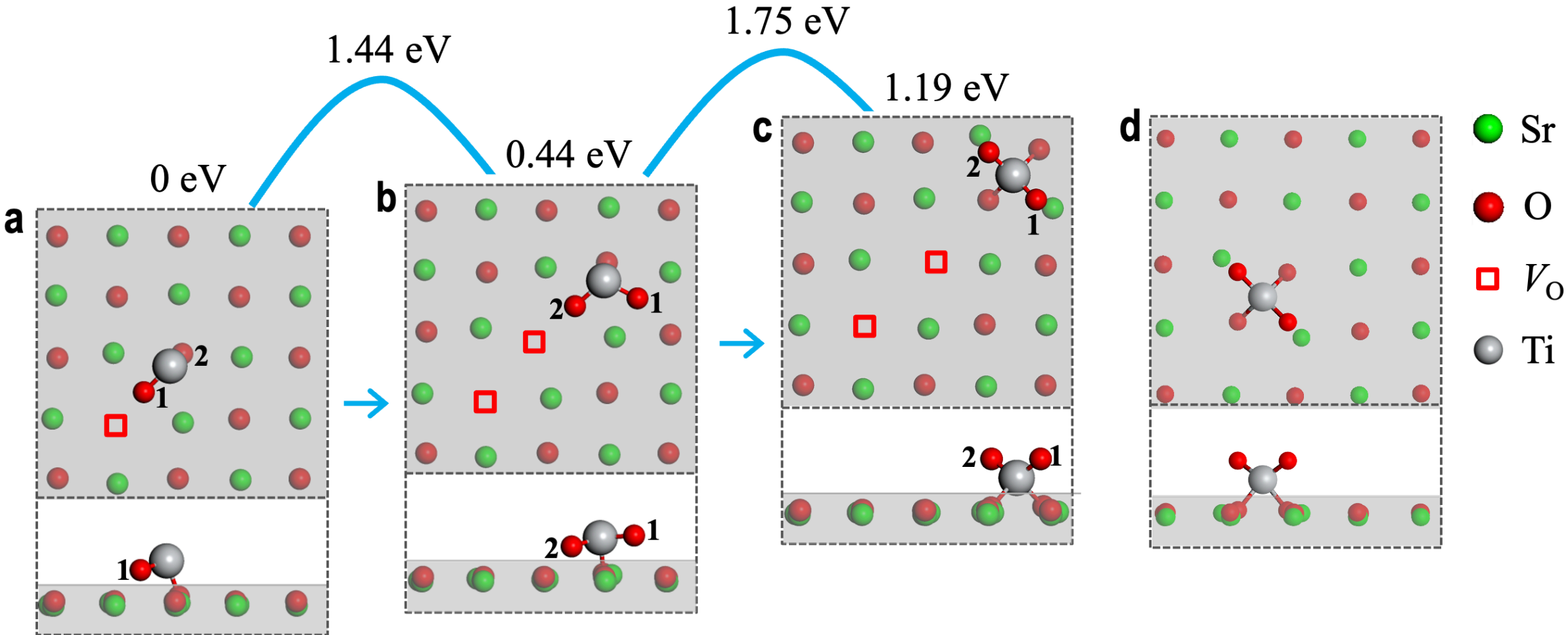

**FIG. 12.** Reactions of $Ti^*$ on the SrO-terminated surface. Several atoms are numbered for clarity.

Similar to the SrO-terminated surface, $Ti^*$ on the bilayer SrO-terminated surface can form a $TiO_2^*$ molecule [Fig. 13(a)] and rotate to produce $TiO_3^*$ [Fig. 13(b)] and $TiO_4^*$ [Fig. 13(c)] by extracting oxygen from the surface. Both the $TiO_3^*$ and $TiO_4^*$ molecules are energetically more stable than the $TiO_2^*$ molecule by 0.78 and 1.08 eV, respectively. The reaction barriers for the $TiO_2^* \rightarrow TiO_3^*$ and $TiO_3^* \rightarrow TiO_4^*$ transitions are 1.62 and 0.91 eV, respectively. At the growth temperature of 1023 K, these processes can occur on a timescale of $\sim 10^{-5}$ s per event, which is noticeably shorter than the growth time of each atomic layer.

In our previous research, we demonstrated that a $TiO_2^*$ molecule on the bilayer SrO-terminated surface can insert into the topmost SrO layer and lift a fragment of the SrO layer, ultimately leading to the growth of a complete $TiO_2$ layer sandwiched between the two SrO layers rather than growing a $TiO_2$ layer above the SrO bilayer [15]. In this study, we observe that Ti* can also permeate an SrO bilayer, with an overall reaction barrier of 1.59 eV (Fig. 13). Given this phenomenon and the observation in our previous study [15], depositing a $TiO_2$ layer on an SrO trilayer is necessary to achieve a desired $TiO_2$–SrO–SrO or $Sr_2TiO_4$ stacking.

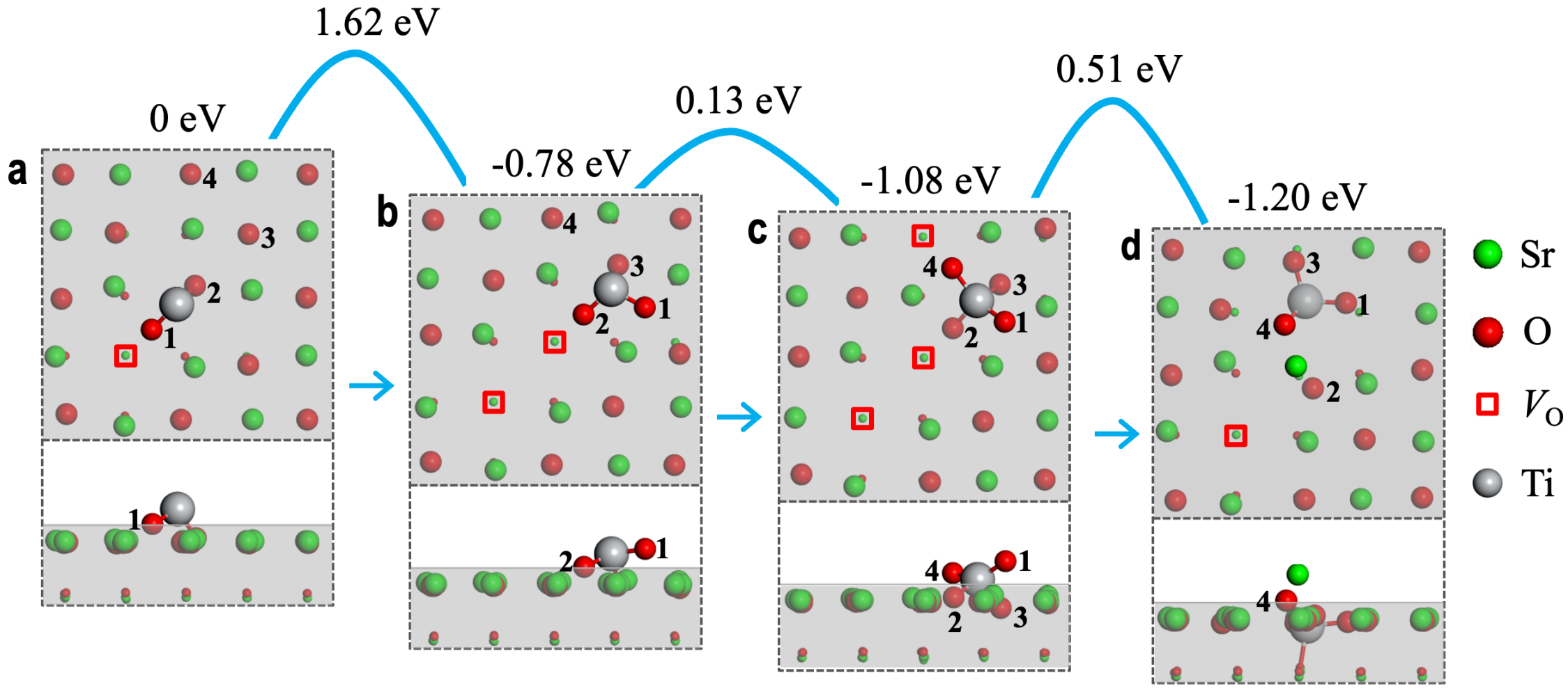

**FIG. 13.** Reactions of Ti* on the bilayer SrO-terminated surface. Several atoms are numbered for clarity.

## IV. CONCLUSIONS

We computationally investigate the atomistic processes involved in the MBE growth of $SrTiO_3$ and $Sr_2TiO_4$ films on $SrTiO_3$ substrate using first-principles thermodynamics, transition-state calculations, and AIMD simulations. Our findings reveal that the molecular beams of Sr and Ti predominantly consist of single atoms. In the epitaxial growth of $SrTiO_3$, the diffusions of Sr* and SrO* on the $TiO_2$-terminated surface involve the dynamic generation of oxygen vacancies in the $TiO_2$ layer. The relatively low diffusion barriers and the absence of an E-S barrier for $SrO_x$ species suggest a 2D layer growth for the SrO layer, while the presence of unintentional $Ti_{Sr}$ defects could promote island growth. $TiO_x$ molecules, on the other hand, do not easily diffuse on the SrO-terminated surface due to their strong binding with the surface, which is anticipated to inhibit island formation. In the growth of $Sr_2TiO_4$, the diffusion of SrO* on the SrO-terminated surface also involves the dynamic formation of oxygen vacancies in the SrO layer. Additionally, the $TiO_x$ molecule can insert into the bilayer SrO-terminated surface, necessitating the deposition of a triple SrO layer prior to the deposition of a $TiO_2$ layer during the growth of $Sr_2TiO_4$ film.

We note that previous experimenters have reported $SrTiO_3$ substrates obtained under certain treatments to exhibit a double-$TiO_2$ termination [34–38], where the surfaces display diverse reconstruction patterns. Epitaxial growth on such surfaces is expected to be more complex and would require an integrated use of thermodynamic calculations, molecular dynamics simulations, and transition-state analyses, as demonstrated in this work, to elucidate the underlying mechanisms. Nevertheless, the results presented in this study may also provide insights there when the top surface is not highly defective.

## ACKNOWLEDGMENTS

Theoretical calculations were majorly carried out on the Carbon High-Performance Computing Cluster at Argonne National Laboratory under Rewards No. CNM29783 and No. CNM35702 and on clusters in the National Energy Research Scientific Computing Center, which is supported by the Office of Science of the U.S. Department of Energy under Contract No. DE-AC02-05CH11231. Partial calculations were carried out on the Taiyi cluster supported by Center for Computational Science and Engineering of Southern University of Science and Technology. G.L. gratefully acknowledges partial support of his part of this research by NSF through the University of Wisconsin Materials Research Science and Engineering Center (No. DMR-1121288) and later by the National Foundation of Natural Science, China (No. 52273226). D.M. gratefully acknowledges full support of his part of this research by NSF through the University of Wisconsin Materials Research Science and Engineering Center (No. DMR-1121288 and No. DMR-2309000).

## DATA AVAILABILITY

All key input and output files used in this study, which are permitted for public sharing, have been organized and made available through Figshare [26].

# Supplemental Material for
# Atomistic Modeling of Molecular Beam Epitaxy Growth of $SrTiO_3$ and $Sr_2TiO_4$ Thin Films

Guangfu Luo[1,2,3*] and Dane Morgan[1*]

[1]Department of Materials Science and Engineering, University of Wisconsin-Madison, Wisconsin 53706, USA

[2]State Key Laboratory of Quantum Functional Materials, Department of Materials Science and Engineering, Southern University of Science and Technology, Shenzhen 518055, China

[3]Institute of Innovative Materials, Southern University of Science and Technology, Shenzhen 518055, China

[*]E-mail: (G.L.) luogf@sustech.edu.cn, (D.M.) ddmorgan@wisc.edu

**Table S1.** Growth temperature, molecular beam conditions, and chemical potentials under experimental conditions.[1]

| | Condition | Chemical potential (eV) |
|---|---|---|
| Substrate | $T$ ~1023 K | – |
| Oxygen | $P_{O_2}$ ~1.33 × $10^{-4}$ Pa | -13.91 |
| Sr beam | $P_{Sr}$ ~1.32 × $10^{-2}$ Pa<br>$T$ ~673 K | -2.11 (-2.36†) |
| Ti beam | $P_{Ti}$ ~3.09 × $10^{-2}$ Pa<br>$T$ ~1773 K | -8.17 (-8.83†) |

†Chemical potentials near the substrate, estimated as follows by accounting for molecular-beam expansion between the effusion-cell nozzle and the substrate. Assuming a nozzle-to-substrate distance of $l$ ~20 cm,[2] a nozzle radius of $d$ ~4 mm,[3] and a molecular-beam full divergence angle of $\theta$ ~10°,[4] the volumetric expansion ratio is estimated as $(\theta l)^2/d^2 \approx 76$. Thus, $P_{source}/P_{substrate} \approx 76$, where $P_{source}$ and $P_{substrate}$ denote the partial pressures at the metal source and above the substrate, respectively. The corresponding correction to the chemical potential is therefore $-k_B T \ln(P_{source}/P_{substrate})$, yielding -0.25 eV for the Sr beam at 673 K and -0.66 eV for the Ti beam at 1773 K.

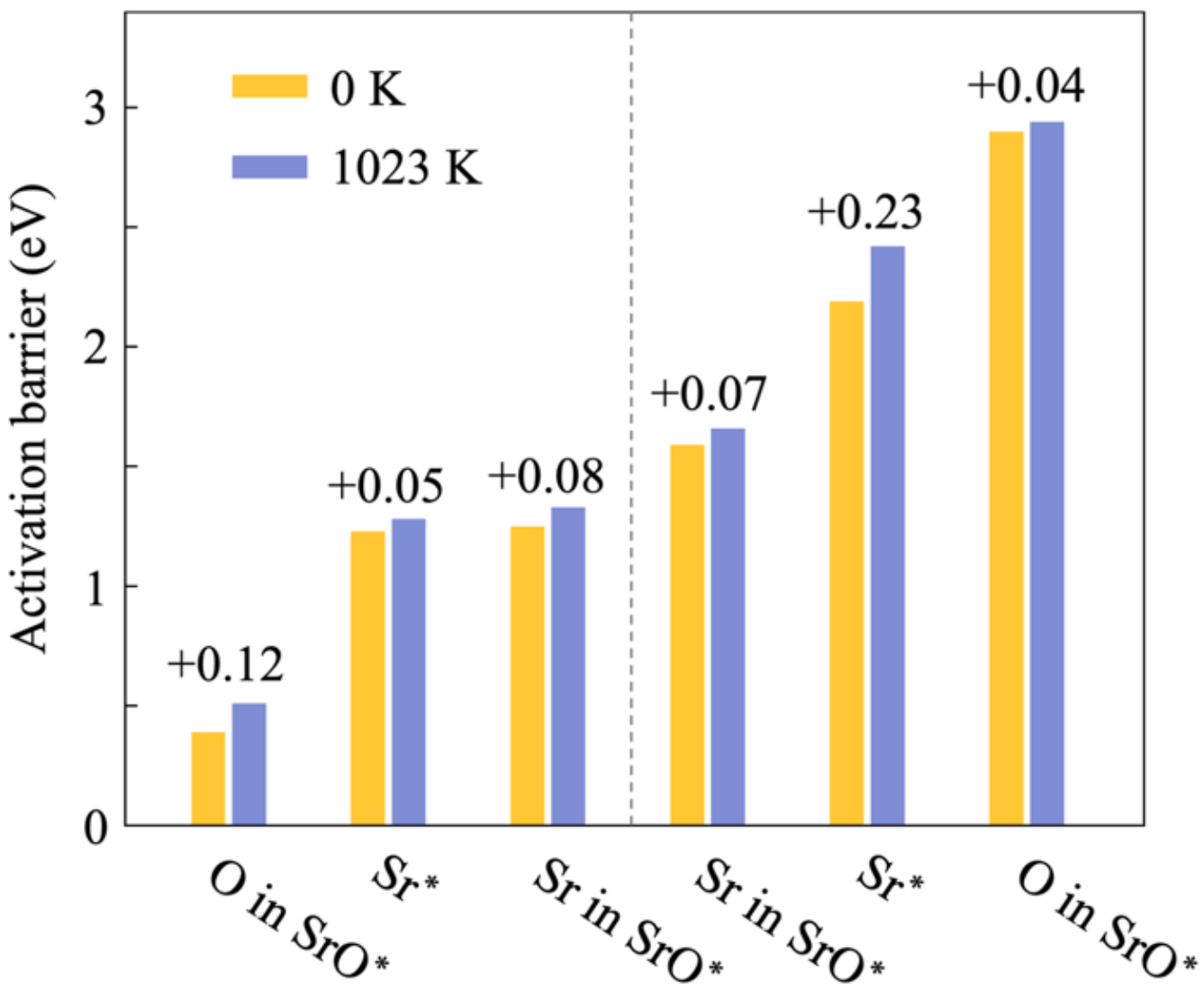

**Figure S1.** Temperatures effects on the hopping barriers of Sr* and SrO* on the $TiO_2$-terminated surface for mechanisms involving $V_O$ in the $TiO_2$ plane (left of the dashed line) and mechanisms not involving $V_O$ (right of the dashed line). The values above the bars indicate the increases of activation barrier at 1023 K relative to those at 0 K, where no thermal corrections are included.

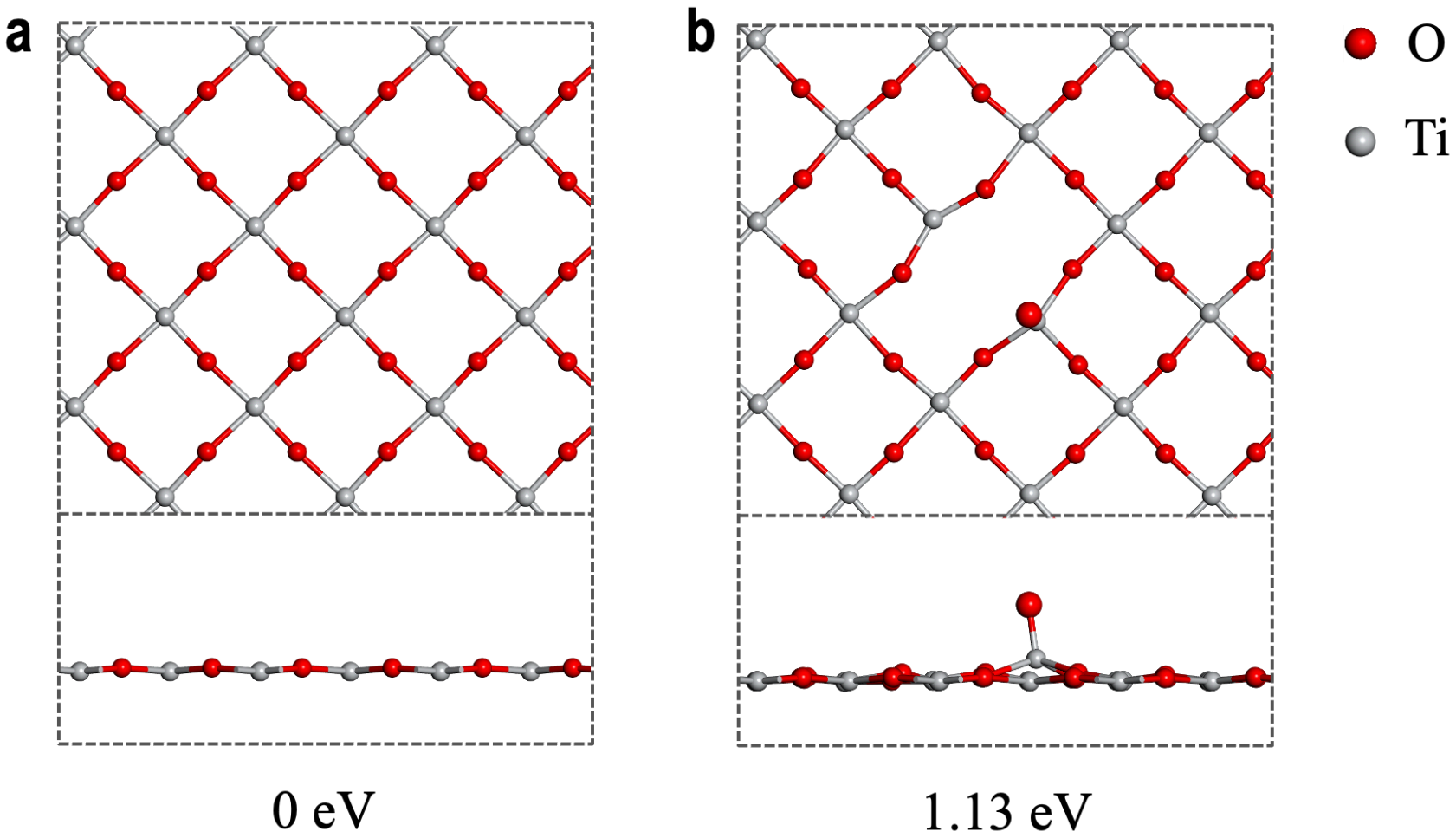

**Figure S2.** (a) Ideal $TiO_2$-terminated surface and (b) $TiO_2$-terminated surface with one oxygen atom displaced to a nearby site analogous to those in the hopping of Sr* and SrO* involving oxygen-vacancy formation. The horizontal coordinates of the displaced oxygen atom are fixed at positions similar to those in the hopping processes, while its vertical coordinate is fully relaxed. The relative energy of the structure with an oxygen vacancy is 1.13 eV, significantly greater than the corresponding value of -0.01 eV for the structure with Sr* adsorption (Fig. 4b). This indicates that Sr* adsorption can markedly promote the formation of oxygen vacancy.